\documentclass[fleqn,usenatbib]{mnras}

\usepackage{graphicx}	
\usepackage{amsmath}	
\usepackage{amssymb}
\usepackage[T1]{fontenc}
\usepackage{etoolbox}
\robustify{\cite}

\usepackage{newtxtext,newtxmath}
\title[Impact of spin on Poynting-Robertson drag]{Impact of neutron star spin on Poynting-Robertson drag during a Type I X-ray burst}

\graphicspath{{./}{figures/}}

\author[J. Speicher et al.]{
J. Speicher$^{1}$\thanks{E-mail: jspeicher3@gatech.edu}\thanks{NDSEG Fellow},
P. C. Fragile$^{2}$ and
D. R. Ballantyne$^{1}$
\\
$^{1}$Center for Relativistic Astrophysics, School of Physics, Georgia
  Institute of Technology, 837 State Street, Atlanta, GA 30332-0430, USA\\
$^{2}$Department of Physics \& Astronomy, College of Charleston, 66 George St., Charleston, SC 29424, USA
}
% ...

\date{Accepted XXX. Received YYY; in original form ZZZ}

\pubyear{???}

\begin{document}
\label{firstpage}
\pagerange{\pageref{firstpage}--\pageref{lastpage}}
\maketitle

\begin{abstract}
External irradiation of a neutron star (NS) accretion disc induces Poynting-Robertson (PR) drag, removing angular momentum and increasing the mass accretion rate. Recent simulations show PR drag significantly enhancing the mass accretion rate during Type I X-ray bursts, which could explain X-ray spectral features such as an increase in the persistent emission and a soft excess. However, prograde spin of the NS is expected to weaken PR drag, challenging its importance during bursts. Here, we study the effect of spin on PR drag during X-ray bursts. We run four simulations, with two assuming a non-spinning NS and two using a spin parameter of $a_*=0.2$, corresponding to a rotation frequency of 500 Hz. For each scenario, we simulate the disc evolution subject to an X-ray burst and compare it to the evolution found with no burst. PR drag drains the inner disc region during a burst, moving the inner disc radius outward by $\approx1.6$ km in the $a_*=0$ and by $\approx2.2$ km in the $a_*=0.2$ simulation. The burst enhances the mass accretion rate across the innermost stable circular orbit $\approx7.9$ times when the NS is not spinning and $\approx11.2$ times when it is spinning. The explanation for this seemingly contradictory result is that the disc is closer to the NS when $a_*=0.2$, and the resulting stronger irradiating flux offsets the weakening effect of spin on the PR drag. Hence, PR drag remains a viable explanation for the increased persistent emission and soft excess observed during X-ray bursts in spinning NS systems.
\end{abstract}


\section{Introduction } \label{sec:intro}
Compact objects such as black holes, neutron stars, and white dwarves may accrete matter from their surroundings. If the infalling material possesses angular momentum, an accretion disc forms to transport the angular momentum outwards through viscous processes \citep[for a review, see][]{Abramowicz2013}. The disc structure is often described by the $\alpha$ model \citep{Shakura1973}. The Newtonian model by \cite{Shakura1973} and its general relativistic counterpart by \cite{Novikov1973} assume no external irradiation. Despite external irradiation coming from the corona, a region of hot electrons \citep[e.g.,][]{Galeev1979,Zdziarski2004}, and the neutron star surface or a boundary layer \citep[e.g.,][]{Inogamov1999,Popham2001,Inogamov2010}, these disc models offer a satisfactory description of realistic accretion discs and are often the basis of accretion disc simulations \citep[e.g.,][]{Kubota2004,Kubota2005,Shafee2008,Fragile2020}. However, when accretion discs are strongly irradiated by Type I X-ray bursts, the disc structure starts to diverge from the standard model.


Type I X-ray bursts can arise from neutron stars in low-mass X-ray binaries \citep[for a review, see][]{Lewin1993, Galloway2021}. In these binaries, neutron stars accrete matter via Roche-Lobe overflow. The accreted material, mainly hydrogen and helium, accumulates on the neutron star's surface. The matter is gradually compressed, increasing the accreted layer's temperature until nuclear burning sets in. However, due to the thin-shell instability, the layer does not cool by expansion but heats up further \citep[e.g.,][]{Schwarzschild1965, Fujimoto1981, Yoon2004}. Subsequent accretion eventually renders the nuclear burning unstable, and the accreted matter conflagrates within seconds, leading to an X-ray burst \citep[e.g.,][]{Fujimoto1981, Narayan2003, Galloway2021}. During its fast exponential rise, followed by a slow exponential decay, the burst emits $\sim 10^{39}-10^{42}$ erg of radiation into the neutron star environment \citep[e.g.,][]{Strohmayer2002, Kuulkers2003, Galloway2008, Degenaar2013}. 

The burst radiation observably impacts the accretion environment \citep[for a review, see][]{Degenaar2018}. Observations and calculations find reduced high energy emission during bursts, which has been linked to Compton cooling of electrons in the corona \citep[e.g.,][]{Maccarone2003, Ji2014, Sanchez2020, Speicher2020}. The accretion disc's reflection spectrum also evolves over the burst's duration. Several burst observations feature emission lines \citep[e.g.,][]{Degenaar2013, Keek2014, Keek2017, Bult2019}, whose strength should evolve with the burst flux \citep[e.g.,][]{Speicher2022}. Observed evolutions in reflection spectra have been limited to superbursts, which have been used to deduce a change in disc ionization and inner accretion disc radius \citep[e.g.,][]{Ballantyne2004, Keek2014APJL}.

In addition, the overall amount of persistent emission changes during bursts. \cite{Worpel2013} and \cite{Worpel2015} multiplied the persistent emission by a normalization factor, $f_a$, and not only found that including $f_a$ improves the fit, but also that $f_a$ often exceeded 1. Due to its ability to improve fits, $f_a$ has become a common burst analysis tool \citep[e.g.,][]{Jaisawal2019,Bult2022, Guver2022MNRAS, Guver2022}. Furthermore, the normalization factor has been used to accommodate a soft excess occurring at energies below 3 keV 
\citep[e.g.,][]{Keek2018, Bult2019, Guver2022MNRAS, Guver2022}. The enhanced normalization factors have been linked to an increased mass accretion rate, $\dot{M}$, caused by burst photons exerting Poynting-Robertson (PR) drag on the accretion disc. PR drag is a special relativitic effect, through which irradiation from a central source removes angular momentum from an orbiting particle \citep[][]{Robertson1937}. As the calculations by \cite{Walker1989} and \cite{Walker1992} showed, the additional irradiation by the burst photons causes particles to fall faster toward the neutron star, thus increasing $\dot{M}$.

The effect of PR drag during an X-ray burst was studied in simulations by \cite{Fragile2020}. In their simulations, an X-ray burst with a peak luminosity of $L_{0}=10^{38}$ erg s\textsuperscript{-1} interacted with a thin accretion disc. The increase in PR drag increased $\dot{M}$ by over an order of magnitude. The increased mass accretion rate drained the inner accretion disc region and the inner disc radius moved outward by a few kilometers. 

However, the impact of PR drag measured by \cite{Fragile2020} was under the assumption of a nonspinning neutron star, which may have made the PR drag more pronounced than in real neutron star systems. Realistically, neutron stars spin with frequencies up to $\sim700$ Hz \citep[e.g.,][]{Chakrabarty2003, Hessels2006, Bassa2017, Patruno2017}. This usually prograde neutron star spin will reduce the angular velocity difference between the neutron star photon field and the gas of the disc. Hence, a prograde neutron star spin is expected to reduce the strength of PR drag and thus the mass accretion rate enhancement \citep[e.g.,][]{Walker1992}. Since the mass accretion rate enhancement has served as an explanation for an increased persistent emission or soft excess during burst observations, it is worth reassessing whether PR drag is strong enough in spinning neutron star systems to enhance $\dot{M}$ significantly, or whether an increase in the $f_a$ factor is driven by other processes, such as reflection \citep[e.g.,][]{Speicher2022}.

Hence, this paper studies the importance of PR drag during an X-ray burst for a spinning neutron star, for which we extend the simulations by \cite{Fragile2020}. Unless otherwise specified, all equations in this paper are written in units where $G=c=1$. The paper is structured as follows: Sec. \ref{sec:methods} outlines the methods relevant to our simulations. The simulation results will be presented in Sec. \ref{sec:results} with additional analytical calculations of PR drag provided in Sec.~\ref{subsec:angularForce}. Finally, we discuss the implications of our results in Sec. \ref{sec:discussion} with conclusions stated in Sec. \ref{sec:conclusion}. 
\section{Methods}\label{sec:methods}

Our simulations extend the work by \cite{Fragile2020}, which implemented an X-ray burst interacting with a thin accretion disc around a non-spinning neutron star. The neutron star spin we introduce corresponds to a rotational frequency, $f_*$, in the prograde direction. To highlight the effect of the neutron star spin, we select a high, yet realistic, rotational frequency of $500$ Hz. 
The dimensionless spin parameter, $a_*$, is defined as
\begin{equation}
    a_* = \frac{J}{M^2}, \label{eq:spinParam}
\end{equation}
where $J$ is the angular momentum and $M$ is the neutron star mass. We assume that $J$ is the product of the moment of inertia and the angular frequency ($2\pi f_*$). The relationship between the rotational frequency and the spin parameter depends on the neutron star's equation of state \citep[EOS; e.g.,][]{Sibgatullin1998, Koliogiannis2020}. We select EOS A by \cite{Cook1994}, because its compactness best matches our assumed neutron star parameters. For our simulations, we choose a neutron star radius of $R_*=10.7$ km and mass $M=1.45$ M$_\odot$. Using the tabulated neutron star properties of EOS A, which relates angular frequencies, moments of inertia, and neutron star masses \citep[see table 9 by][]{Cook1994}, we follow \cite{Sibgatullin1998} and calculate an array of spin parameters and corresponding angular velocity parameters $W$,
\begin{equation}
    W=2\pi f_* M. \label{eq:angularVParam}
\end{equation}
From this spin versus angular velocity parameter relationship, we can infer that the spin parameter corresponding to $f_*=500$ Hz and $M=1.45$ M$_\odot$ is $a_*\approx0.2$.

The neutron star rotation impacts the four-momentum and stress-energy tensor of the burst photons. The photon's four-momentum is, in Boyer-Lindquist coordinates $\{t,r,\theta,\phi\}$ \citep[e.g.,][]{Bakala2019},

\begin{eqnarray}
k^t & = & \Sigma^{-1} \left[ab - a^2 \sin^2 \theta + \left(r^2 + a^2\right) P \Delta^{-1}\right] \nonumber \\
k^r & = & \Sigma^{-1} \sqrt{R_{b,q}(r)} \nonumber \\
k^\theta & = & \pm \Sigma^{-1} \sqrt{\Theta_{b,q}(\theta)} \nonumber \\
k^\phi & = & \Sigma^{-1} (b \csc^2 \theta - a + a P \Delta^{-1}) ~,\label{eq:4momentum}
\end{eqnarray}
where $P \equiv r^2 + a^2 -ba$, $\Delta \equiv r^2 - 2Mr + a^2$, and $\Sigma \equiv r^2 + a^2 \cos^2 \theta$. The terms $\Theta_{b,q}(\theta)$ and $R_{b,q}(r)$ are $\Theta_{b,q}(\theta) \equiv q + a^2 \cos^2 \theta -b^2 \cot^2 \theta$ and $R_{b,q}(r) = (r^2 + a^2 - ab)^2 - \Delta [q+(b-a)^2]$, respectively. Here, $a$ is the specific angular momentum of the neutron star ($J/M$). The $\pm$ sign in the $k^\theta$ expression (eq. \ref{eq:4momentum}) depends on the photon trajectory. The latitudinal photon impact parameter, $q$, is given as
\begin{align}
q &\equiv \left(\frac{k_\theta}{k_t}\right)^2 + \left[b \tan\left(\frac{\pi}{2}-\theta\right)\right]^2 - a^2 \cos^2 \theta \nonumber \\
 &= b^2 \cot^2 \theta_e - a^2 \cos^2 \theta_e,
\end{align}
while the azimuthal impact parameter, $b$, is
\begin{equation}
    b \equiv -\frac{k_\phi}{k_t} = \frac{\sin^2\theta_e [(R_*^4 + a^2\mathcal{B})2\pi f_* - 2aMR_*]}{R_*^2 - 2MR_* + a(a \cos^2 \theta_e + 2MR_*2 \pi f_* \sin^2 \theta_e)},
\end{equation}
where $\theta_e$ is the polar emission angle. The term $\mathcal{B}$ is given as 
\begin{equation}
    \mathcal{B} = (a^2 + R_*^2) \cos^2 \theta_e + 2MR_* \sin^2 \theta_e + R_*^2.
\end{equation}

To match the space-time metric of the simulation, the photon momentum components are transformed from Boyer-Lindquist to Kerr-Schild coordinates. The photon 4-momentum (eq. \ref{eq:4momentum}) transforms as 
\begin{eqnarray}
k^{\tilde{t}} & = & k^t + \frac{2Mr}{\Delta} k^r \nonumber \\
k^{\tilde{r}} & = & k^r \nonumber \\
k^{\tilde{\theta}} & = & k^\theta \nonumber \\
k^{\tilde{\phi}} & = & k^\phi + \frac{a}{\Delta}k^r ~. \label{eq:photon4Transform}
\end{eqnarray}

Since the simulation mesh is more refined at smaller radii and towards the equatorial plane \citep{Fragile2020}, the components $k^{\tilde{r}}$ and $k^{\tilde{\theta}}$ are additionally converted as 
\begin{eqnarray}
k^{\tilde{x}_1} & = & \left(\frac{\partial \tilde{r}}{\partial \tilde{x}_1}\right)^{-1} k^{\tilde{r}} \nonumber \\
k^{\tilde{x}_2} & = & \left(\frac{\partial \tilde{\theta}}{\partial \tilde{x}_2}\right)^{-1} k^{\tilde{\theta}} ~,
\end{eqnarray}
where 
\begin{eqnarray}
\tilde{r} &= R_* \exp(\tilde{x}_1 - 1)\nonumber\\
\tilde{\theta} &= \tilde{x}_2 + 0.25 \sin \tilde{x}_2.\label{eq:meshCoordinates}
\end{eqnarray}
The coordinate $\tilde{x}_2$ describes the uniform latitude coordinate in the code.

\begin{figure}
    \centering
    \includegraphics[width  =0.45\textwidth]{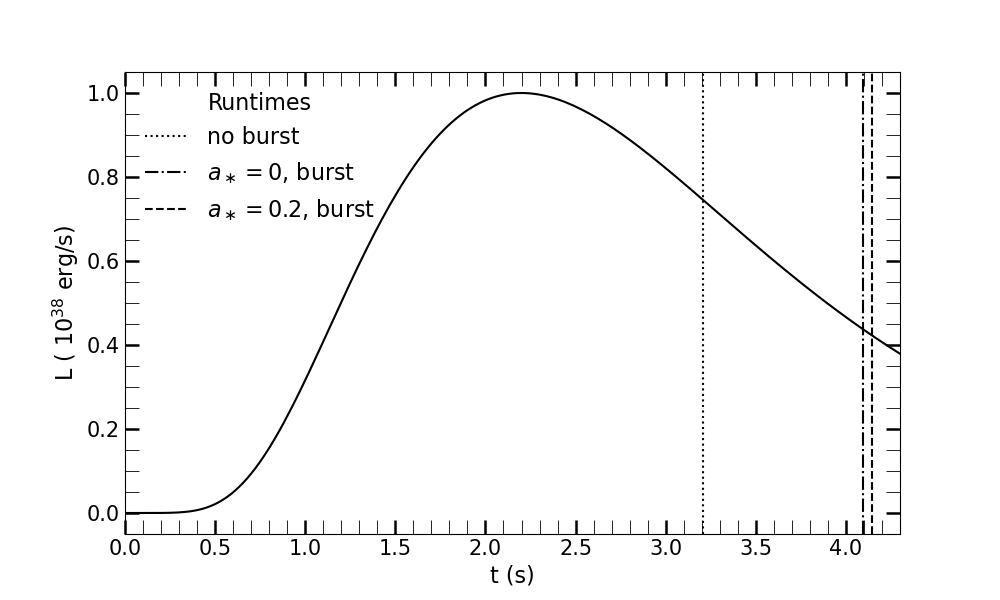}
    \caption{Burst profile (eq. \ref{eq:luminosity}). The vertical lines mark the runtimes of the burst (dot-dashed line for $a_*=0$ and dashed line for $a_*=0.2$) and no-burst (dotted line) simulations.}
    \label{fig:luminosity}
\end{figure}

The photon momentum enters the photon stress-energy tensor, $R^{\mu\nu}$, as \citep[e.g.,][]{Bakala2019} 
\begin{equation}
    R^{\tilde{\mu}\tilde{\nu}} = \Phi^2 k^{\tilde{\mu}} k^{\tilde{\nu}},\label{eq:stressEnergyTensor}
\end{equation}
where the intensity parameter, $\Phi$, is normalized so as to produce the desired integrated burst luminosity
\begin{equation}
    L = -\int_S \sqrt{-g} R^{\tilde{r}}_{\tilde{t}} dA_{\tilde{r}} ~,
\end{equation}
where $dA_{\tilde{r}}$ is the surface area element normal to the radial direction and $g$ is the metric determinant.
The luminosity is chosen to vary with time as \citep[Fig. \ref{fig:luminosity}, e.g.,][]{Norris2005,Barriere2015}
\begin{equation}
    L(t) = L_0 e^{2\left ( \tau_1/\tau_2 \right )^{1/2}} e^{-\frac{\tau_1}{t-t_s}-\frac{t-t_\mathrm{s}}{\tau_2}},\label{eq:luminosity}
\end{equation}
where $L_0=10^{38}$ erg s$^{-1}$, $t_\mathrm{s} = -0.25$ s, $\tau_1=6$ s, and $\tau_2=1$ s. Unlike the simulations by \cite{Fragile2020}, we only consider the burst luminosity and ignore any luminosity due to accreted material hitting the neutron star surface.

Similar to the approach by \cite{Sadowski2013}, we use the components $R^{\tilde{r}}_{\tilde{\mu}} = g_{\tilde{\mu}\tilde{\nu}} \Phi^2 k^{\tilde{r}}k^{\tilde{\nu}}$, with $g_{\tilde{\mu}\tilde{\nu}}$ being the metric, to solve the equation 
\begin{equation}
    R^{\tilde{r}}_{\tilde{\mu}} = \frac{4}{3} E_R u^{\tilde{r}}_R (u_R)_{\tilde{\mu}} + \frac{1}{3} E_R \delta^{\tilde{r}}_{\tilde{\mu}}
\end{equation}
for the radiation energy density, $E_R$, in its rest frame and the radiation velocity, $u_R$, in an observer frame. These energy and velocity are used to inject radiation at the inner simulation boundary.

We implemented our simulations in the general relativistic hydrodynamics code \textit{Cosmos++} \citep[][]{Anninos2005,Fragile2012,Fragile2014}. For each simulation, the code first initializes the gas-pressure dominant region of an $\alpha$-disc with a viscosity parameter of $\alpha=0.025$ \citep[e.g.,][]{Abramowicz2013}. The discs start out in hydrostatic and thermal equilibrium. The remaining simulation space is filled with a low energy, low mass density gas with energy $3/2\times 10^{-4}E_{\mathrm{max}}r^{-2.5}$ and density $10^{-6}\rho_{\mathrm{max}}r^{-1.5}$. The terms $E_{\mathrm{max}}$ and $\rho_{\mathrm{max}}$ are the maximum energy and mass density of the disc, respectively. For $a_*=0$, $E_{\mathrm{max}}\approx8.2\times10^{14}$ erg cm\textsuperscript{-3} and $\rho_{\mathrm{max}}\approx0.35$ g cm\textsuperscript{-3}, while for $a_*=0.2$, $E_{\mathrm{max}}\approx1.1\times10^{15}$ erg cm\textsuperscript{-3} and $\rho_{\mathrm{max}}\approx0.43$ g cm\textsuperscript{-3}. During the simulation, \textit{Cosmos++} numerically advances the general relativistic, radiative, viscous hydrodynamics equations as outlined by \cite{Fragile2018}.  The simulations are two-dimensional and cover a radial range of 10.7 km $\lesssim r\lesssim$ 352 km and $0\lesssim\theta\lesssim\pi$, with the innermost zones lying just above the stellar surface (i.e., the neutron star surface is not in the simulation domain and always within the innermost stable circular orbit of the disc). The simulation domain is covered by a mesh subdivided into the equivalent of 384 cells in the radial direction, and 384 cells in the angular direction. The exponential cell spacing in the radial direction and a greater concentration of cells in the equatorial plane using $\tilde{x}_1$ and $\tilde{x}_2$ (eq. \ref{eq:meshCoordinates}) allows a greater mesh refinement closer to the neutron star and the midplane of the disc.  All simulations omit the effects of the stellar magnetic field. That is, we assume the field is weak enough to not impact the dynamics of the accretion flow.

As a change to the underlying physics described in \cite{Fragile2020}, we use Planck and Rosseland mean opacity coefficients of $3.2\times10^{24}$ cm\textsuperscript{2} g\textsuperscript{-1} and $8.6\times10^{22}$ cm\textsuperscript{2} g\textsuperscript{-1}, respectively \citep{Hirose2009}. To study the impact of spin, we ran two sets of simulations, one with spin and one without. Each set consists of a burst and a no-burst simulation. The no-burst simulations (Fig. \ref{fig:luminosity}, vertical dotted line) ran for $\approx3.2$ s, while the burst simulations (vertical dot-dashed line for $a_*=0$ and vertical dashed line for $a_*=0.2$) ran for $\approx4.1$ s. 
\section{Results}\label{sec:results}

\begin{figure*}
     \centering
    \includegraphics[width = 0.8 \textwidth]{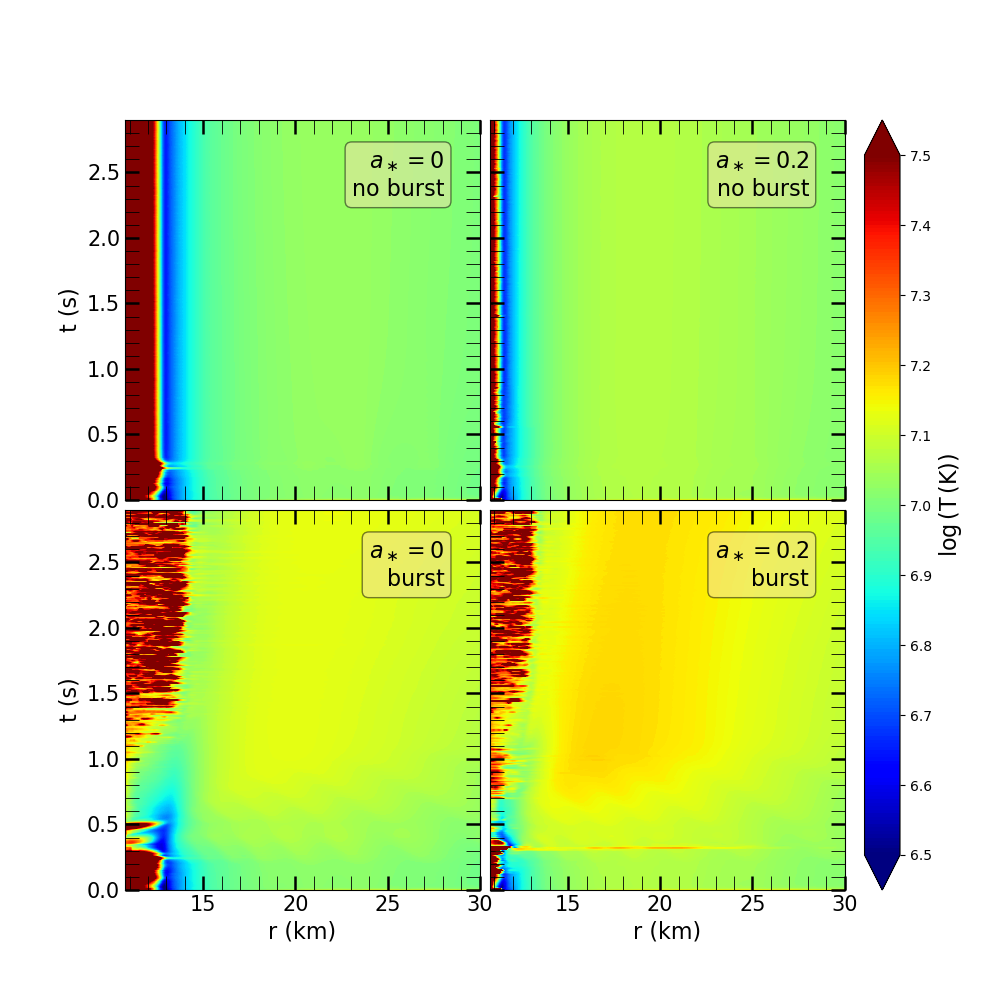}
    \caption{Spacetime diagrams of the density-weighted temperature. Top row: The temperatures remain constant in the no-burst simulations. The spin changes the spacetime and thus the accretion disc structure, moving $r_{\mathrm{isco}}$ closer to the neutron star in the $a_*=0.2$ simulation (top row, right panel) than in the $a_*=0$ one (top row, left panel). Bottom row: Due to the burst radiation, the inner disc radii recede outwards for both the $a_*=0$ (bottom row, left panel) and the $a_*=0.2$ (bottom row, right panel) simulations. Additionaly, the burst radiation heats the discs. The final disc temperatures differ between the spin and the no-spin simulation due to the different steady-state disc structures (top row).}
    \label{fig:tempAll}   
\end{figure*}

Fig. \ref{fig:tempAll} shows spacetime diagrams of the logarithm of the density-weighted gas temperature for all four simulations. In the absence of a burst, the simulated accretion discs are constant in time (upper row). Differences between these undisturbed discs arise through the impact of the neutron star spin on the spacetime. The accretion discs in the no-burst simulations terminate at the innermost stable circular orbit $r_{\mathrm{isco}}$, which is spin dependent. For $a_*=0$ (upper row, left panel), $r_{\mathrm{isco}} \approx12.85$ km, and for $a_*=0.2$ (upper row, right panel), $r_{\mathrm{isco}} \approx11.41$ km.
Besides the differences in $r_\mathrm{isco}$, in the no-burst simulations the $a_*=0.2$ disc (upper row, right panel) is hotter than the $a_*=0$ one (upper row, left panel) at radii larger than the $r_\mathrm{isco}$ of $a_*=0$. 
In the burst simulations (Fig. \ref{fig:tempAll}, lower row), the burst radiation heats the disc. Due to its higher steady-state temperatures (upper row, right panel), the simulated $a_*=0.2$ disc reaches higher temperatures at radii $\gtrsim14$ km compared to the $a_*=0$ disc (lower row, left panel) during the burst. The temperature increase inflates both discs since they are pressure supported. 

\begin{figure}
    \centering
    \includegraphics[width = 0.45\textwidth]{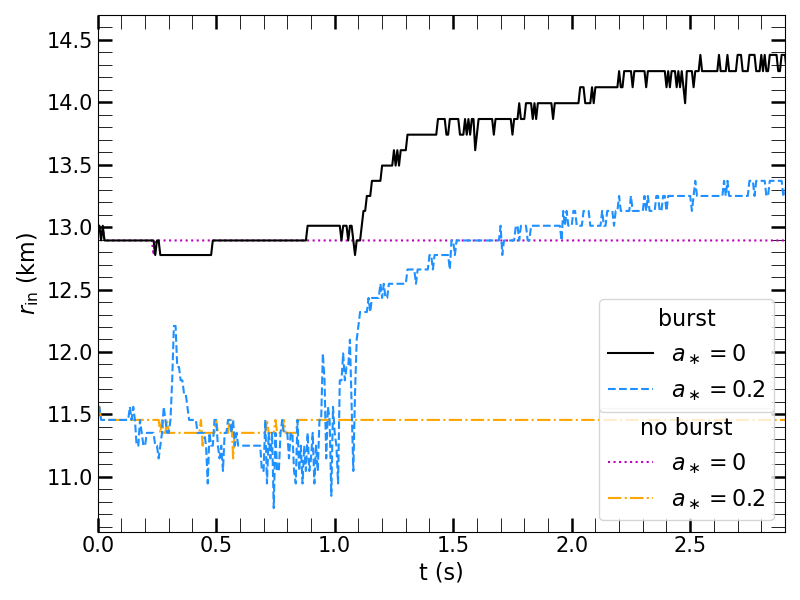}
    \caption{Movement of the inner accretion disc radius. The inner radius is set at a radius where the surface density is closest to $2.5\times10^2$ g cm\textsuperscript{-2}. While $r_{\mathrm{in}}$ remains at $\approx r_{\mathrm{isco}}$ in the $a_*=0$ (magenta dotted line) and $a_*=0.2$ (orange dot-dashed line) no-burst simulations, $r_{\mathrm{in}}$ recedes outward by $\approx1.6$ km in the $a_*=0$ (black solid line) and by $\approx2.2$ km in the $a_*=0.2$ (blue dashed line) burst simulations.}
    \label{fig:rIn}
\end{figure}

Another structural change due to the burst is the movement of the inner disc radius, $r_{\mathrm{in}}$, to larger radii (Fig. \ref{fig:rIn}, see also the expansion of the hot optically thin region colored in red in the lower row of Fig. \ref{fig:tempAll}). We define the location of $r_{\mathrm{in}}$ as the radial position where the surface density is closest to $2.5\times10^2$ g cm\textsuperscript{-2}. This surface density value ensures that $r_{\mathrm{in}}\approx r_{\mathrm{isco}}$ in the no-burst simulations (magenta dotted line for $a_*=0$, orange dot-dashed line for $a_*=0.2$). At early times, the burst impact is negligible, and $r_{\mathrm{in}}\approx r_{\mathrm{isco}}$ in all simulations. As the burst luminosity increases (eq. \ref{eq:luminosity}, Fig. \ref{fig:luminosity}), the inner disc radii move outward by $\approx1.6$ km in the $a_*=0$ and by $\approx2.2$ km in the $a_*=0.2$ burst simulation. 

\begin{figure*}
    \centering
    \includegraphics[width = 0.8 \textwidth]{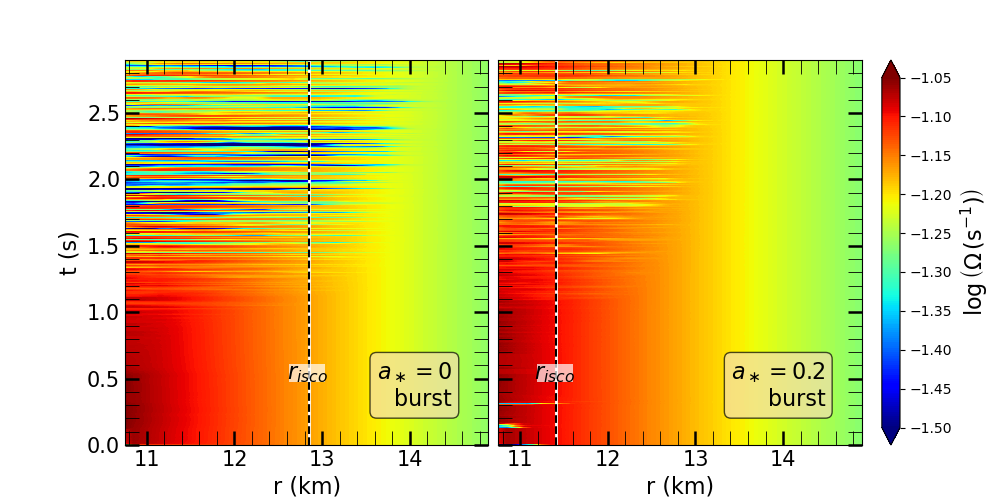}
    \caption{The logarithm of the density-weighted angular frequencies during the burst. As the burst strength increases, the increasing PR drag decreases the angular frequencies in both the $a_*=0$ (left panel) and the $a_*=0.2$ (right panel) burst simulations. Due to the numerical nature of the simulations, the decrease is not smooth but fluctuates. The region of decreased $\Omega$ extends towards larger radii over time as the strengthening burst drains the inner disc region. The dashed vertical lines show the respective $r_{\mathrm{isco}}$, where $\Omega$ is measured in Fig. \ref{fig:omegaRatio}.}
    \label{fig:omegaBurst}
\end{figure*}

\begin{figure}
    \centering
    \includegraphics[width = 0.45 \textwidth]{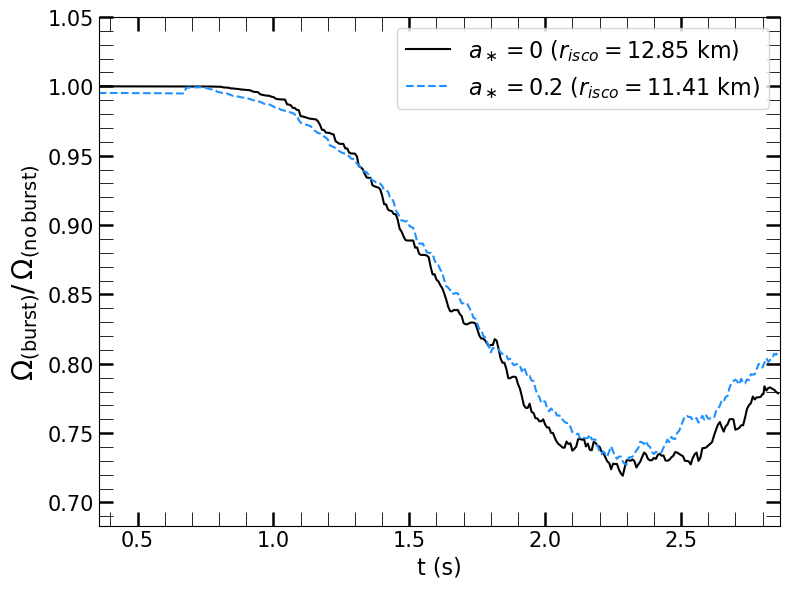}
    \caption{Ratio of the density-weighted angular velocities. The moving averages of the velocities, measured at the respective $r_{\mathrm{isco}}$ (dashed black lines in Fig. \ref{fig:omegaBurst}) depending on the simulation, were calculated with an average window of $\Delta t\approx0.71$ s. The angular frequencies decrease by a similar factor in the $a_*=0$ (black solid line) and the $a_*=0.2$ (blue dashed line) simulations, and reach their minima at about the time the burst peaks.}
    \label{fig:omegaRatio}
\end{figure}

The movement of $r_{\mathrm{in}}$ is due to increased PR drag, which removes angular momentum from the disc. Fig. \ref{fig:omegaBurst} shows the spacetime diagrams of the logarithm of the angular frequency, $\Omega$, for the burst simulations. At $t\lesssim 1$ s, the low burst luminosity does not significantly affect the angular frequencies. Since the disc gas orbits faster closer to the neutron star, the angular frequencies increase with decreasing radii. The increasing burst strength decreases $\Omega$ at $t\gtrsim 1$ s. We find that the decrease in $\Omega$ is not smooth, but fluctuates. The decrease in $\Omega$ can be measured across larger distances with increasing burst luminosity and thus increasing $r_{\mathrm{in}}$ (see the blue and green stripes extending towards larger radii in Fig. \ref{fig:omegaBurst}). 

Even though the evolution of $\Omega$ appears to differ between the $a_*=0$ (Fig. \ref{fig:omegaBurst}, left panel) and $a_*=0.2$ (Fig. \ref{fig:omegaBurst}, right panel) burst simulations, the decrease in angular frequency is comparable at their respective $r_{\mathrm{isco}}$ (dashed black lines in Fig. \ref{fig:omegaBurst}). For Fig. \ref{fig:omegaRatio}, we smoothed out time-dependent fluctuations by calculating the moving averages of $\Omega$ at the respective $r_{\mathrm{isco}}$ with an average window of $\Delta t\approx0.71$ s before taking the ratios between the burst and the no-burst simulation results. Following the $\Omega$ evolution shown in Fig. \ref{fig:omegaBurst}, the ratios of the smoothed frequencies decrease with increasing burst luminosity (eq. \ref{eq:luminosity}) and reach their minima at approximately the time of the burst peak (Fig. \ref{fig:luminosity}). The $\Omega$ ratio reaches a minimum of $\approx0.72$ in the $a_*=0$ simulation and a minimum of $\approx0.73$ in the $a_*=0.2$ simulation at the respective $r_{\mathrm{isco}}$.

\begin{figure*}
    \centering
    \includegraphics[width = 0.8 \textwidth]{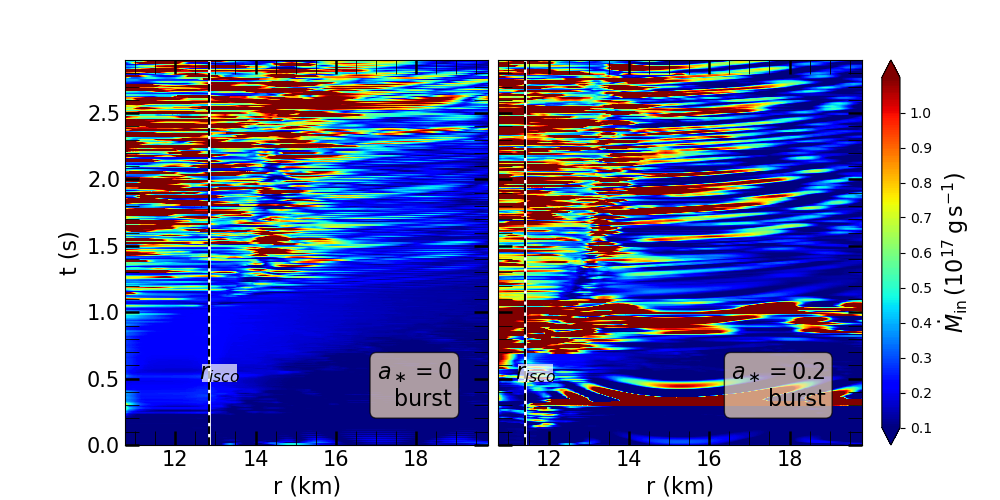}
    \caption{Same as Fig. \ref{fig:omegaBurst}, but showing the mass accretion rate of inflowing material. Initially, $\dot{M}_{\mathrm{in}}\approx 10^{16}$ g s\textsuperscript{-1} across both simulated discs. After $\approx1$ s, $\dot{M}_{\mathrm{in}}$ is significantly enhanced in both the $a_*=0$ (left panel) and the $a_*=0.2$ (right panel) burst simulations. The spike in $\dot{M}_{\mathrm{in}}$ at $t\sim0.5-1$ s in the $a_*=0.2$ simulation is a transient numerical effect. The enhanced $\dot{M}_{\mathrm{in}}$ after $t\sim1$ s extends towards larger radii over time as the burst strengthens and the $\dot{M}_{\mathrm{in}}$ enhancement drains the inner disc region.}
    \label{fig:mdotBurst}
\end{figure*}

The loss of angular momentum (Fig. \ref{fig:omegaBurst}, \ref{fig:omegaRatio}) increases the inflowing mass accretion rate $\dot{M}_{\mathrm{in}}$ (Fig. \ref{fig:mdotBurst}). The total mass accretion rate is calculated as 
\begin{equation}
    \dot{M} = 2\pi \int_0^\pi \sqrt{-g}\rho u^r\,d\theta,
\end{equation}
where $u^r$ is the radial 4-velocity component of the gas, and $\rho$ its density. For the spacetime diagram of the inflowing mass accretion rate shown in Fig. \ref{fig:mdotBurst}, we subtract the outflowing mass rate from the total mass accretion rate to isolate the mass flowing towards the neutron star. When the impact of the burst is still negligible, $\dot{M}_{\mathrm{in}}\approx 10^{16}$ g s\textsuperscript{-1} across both simulated discs. Between $t\sim0.5-1$ s, the inflowing mass accretion rate temporarily increases in the $a_*=0.2$ burst simulation (Fig. \ref{fig:mdotBurst}, right panel) due to transient numerical effects. Neglecting these effects, the inflowing mass accretion rate increases with increasing burst luminosity at $t\gtrsim1$ s in both simulations. Similar to the $\Omega$ evolution, the inflowing mass accretion rate does not increase smoothly but fluctuates. Moreover, the $\dot{M}_{\mathrm{in}}$ enhancement appears to be concentrated in two regions of the simulation space at a given instance of time, which is especially visible in the $a_*=0.2$ burst simulation (Fig. \ref{fig:mdotBurst}, right panel). The region of enhanced $\dot{M}_{\mathrm{in}}$ closer to the neutron star is between the star and the accretion disc and the second region starts at the disc. The space of enhanced $\dot{M}_{\mathrm{in}}$ increasingly extends towards larger radii as the burst strengthens and the $\dot{M}_{\mathrm{in}}$ enhancement drains the inner disc region (also visualized by the red regions in Fig. \ref{fig:tempAll}, bottom row).

\begin{figure}
    \centering
    \includegraphics[width = 0.45 \textwidth]{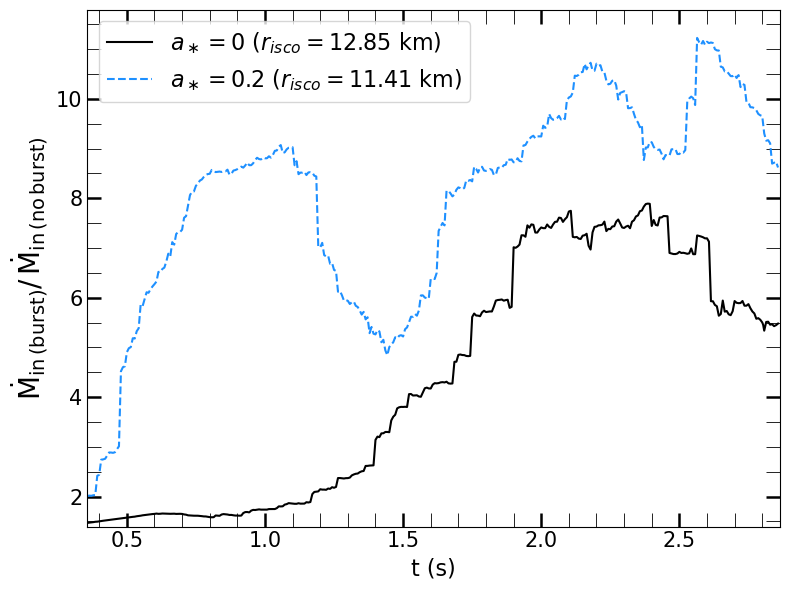}
    \caption{Same as Fig. \ref{fig:omegaRatio}, but showing the inflowing mass accretion rate ratios between burst and no-burst simulations. The inflowing mass accretion rate increases during the burst for both the no-spin (solid, black line) and the spin (blue, dashed line) simulations at their respective $r_{\mathrm{isco}}$ (dashed vertical lines in Fig. \ref{fig:mdotBurst}). The increase in $\dot{M}_{\mathrm{in}}$ for $a_*=0.2$ at t$\sim$1 s is due to transient numerical effects.}
    \label{fig:mdotTimeR}
\end{figure}

Even though a prograde neutron star spin should weaken the PR drag, the inflowing mass accretion rate at the $r_{\mathrm{isco}}$ (black dashed vertical lines in Fig. \ref{fig:mdotBurst}) is actually higher in the $a_*=0.2$ simulation than in the $a_*=0$ one (Fig. \ref{fig:mdotTimeR}). 
The strengthening burst increases PR drag and raises $\dot{M}_{\mathrm{in}}$ compared to the no-burst simulation by a factor of $\approx7.9$ if $a_*=0$ (solid black line) and by $\approx11.2$ if $a_*=0.2$ (blue dashed line) at the respective $r_{\mathrm{isco}}$.

The greater inflowing mass accretion rate enhancement in the $a_*=0.2$ simulation than in the $a_*=0$ one at their respective $r_{\mathrm{isco}}$ (Fig. \ref{fig:mdotTimeR}) is due to different disc structural evolutions. The thin accretion disc equations can be used to approximate $\dot{M}$ as $\left|\dot{M}\right|\sim 3\pi \alpha H^2 \Sigma \Omega$ \citep[see also ][]{Fragile2020}.
The burst radiation heats the discs (Fig. \ref{fig:tempAll}) and, since they are pressure supported, inflates them. 
Besides the increase in height $H$ due to disc inflation, the enhanced $\dot{M}$ due to PR drag lowers the surface density $\Sigma$.
Due to the different steady state disc structures (see e.g. Fig. \ref{fig:tempAll}, upper row), $\Sigma$ and $H$ evolve differently in the spin and the no-spin burst simulation. 
For example, the time-averaged surface densities at the respective $r_{\mathrm{in}}$ (e.g., Fig. \ref{fig:rIn}) decrease during the burst by $\approx24.2\%$ if $a_*=0$ and $\approx17.9\%$ if $a_*=0.2$ compared to the no-burst simulations. Similarly, the time-averaged heights increase due to the burst $\approx4.3$ times if $a_*=0$ and $\approx4.8$ times if $a_*=0.2$ (after its transient event at $t\sim1$s).

\section{Angular force due to PR drag}\label{subsec:angularForce}
The results presented in Section \ref{sec:results} seemingly disagree with the expectation that neutron star spin should reduce PR drag. However, when evaluating the effect of PR drag, one must not only consider its spin-, but also its distance-, dependence. The $a_*=0.2$ disc is closer to the neutron star and therefore experiences a stronger burst flux, which offsets the impact of spin. To illustrate this point, we follow \cite{Klacka1992} and calculate the force due to PR drag as outlined below. 

In the derivation of \cite{Klacka1992}, the unprimed variables are measured in the frame of the irradiating source (the neutron star) and the primed variables in the frame of the moving particle (the disc material). If the primed frame moves with 3-velocity, $\textbf{v}$, variables convert between frames through the usual Lorentz transformations \citep[e.g.,][]{Misner1973} 
\begin{equation}
    x' = \Lambda x,
\end{equation}
where $\Lambda$ is the transformation matrix
\begin{equation}
    \begin{pmatrix}
\gamma & -\gamma v_x & -\gamma v_y & -\gamma v_z\\ 
-\gamma v_x & 1+\left ( \gamma-1 \right ) \frac{v_x^2}{v^2}& \left ( \gamma-1 \right ) \frac{v_xv_y}{v^2} & \left ( \gamma-1 \right ) \frac{v_xv_z}{v^2} \\ 
-\gamma v_y & \left ( \gamma-1 \right ) \frac{v_xv_y}{v^2}  & 1+\left ( \gamma-1 \right ) \frac{v_y^2}{v^2} &  \left ( \gamma-1 \right ) \frac{v_zv_y}{v^2}\\ 
-\gamma v_z &  \left ( \gamma-1 \right ) \frac{v_xv_z}{v^2} &  \left ( \gamma-1 \right ) \frac{v_zv_y}{v^2} &1+\left ( \gamma-1 \right ) \frac{v_z^2}{v^2} 
\end{pmatrix}
\end{equation}
and $\gamma$ is the Lorentz factor $\left(1-\textbf{v}^2\right)^{-1/2}$. The primed vectors convert to the unprimed ones with the inverse of the transformation matrix, $x = \Lambda^{-1} x'$. As an example, the components of the position vector $\left(t,\textbf{x}\right)$ transform as
\begin{align}\begin{split}
    t' &= \gamma \left(t-\textbf{v}\cdot\textbf{x}\right)\\
    \textbf{x}' &= \textbf{x}+\left[\frac{\gamma-1}{\textbf{v}^2}\left(\textbf{v}\cdot\textbf{x}\right)-\gamma t\right]\textbf{v}\end{split}\label{eq:moveP}
\end{align}
and back through
\begin{align}\begin{split}
    t &= \gamma \left(t'+\textbf{v}\cdot\textbf{x}'\right)\\
    \textbf{x} &= \textbf{x}'+\left[\frac{\gamma-1}{\textbf{v}^2}\left(\textbf{v}\cdot\textbf{x}'\right)+\gamma t'\right]\textbf{v}.\end{split}\label{eq:moveUnP}
\end{align}
We assume that the disc particles move on Keplerian orbits with a 3-velocity  \citep[e.g.,][]{Abramowicz2013}
\begin{equation}
    \textbf{v} = \frac{\sqrt{M}}{r^{3/2} + a M^{1/2}}\,\hat{e}_{\phi},
\end{equation}
where $\hat{e}_{\phi}$ is a unit vector in the $\phi$ direction.

In the frame of a disc particle, the incident irradiation energy is 
\begin{equation}
    E_i' = n' A' \epsilon',\label{eq:energyIndisc}
\end{equation}
with $n'$ being the number density of photons, $A'$ the particle cross section and $\epsilon'$ the energy per photon. We assume that the particle's cross section is the Thomson cross section, $\sigma_T$. The incident 3-momentum corresponding to radiation energy $E_i'$ is
\begin{equation}
    \textbf{p}_i' = n' A' \epsilon'\hat{\textbf{S}}',\label{eq:pIndisc}
\end{equation}
where the unit Poynting vector, $\hat{\textbf{S}}$, points in the direction of the irradiation and consists of the normalized stress energy tensor components $R^{0i}$ (eq. \ref{eq:stressEnergyTensor}). For this calculation, we only consider motion in the equatorial plane and set $\theta=\pi/2$. 
We can relate $\hat{\textbf{S}}$ to $\hat{\textbf{S}}'$ by considering the transformation of $(E_i,\textbf{p}_i)$ to the disc particle frame
\begin{equation}\begin{split}
    E_i' &= E_i \gamma \left(1-\textbf{v}\cdot\hat{\textbf{S}}\right)\\
    \textbf{p}_i' &= E_i \left\{\hat{\textbf{S}}+\left[\frac{\gamma-1}{\textbf{v}^2}\left(\textbf{v}\cdot\hat{\textbf{S}}\right)-\gamma \right]\textbf{v}\right\} .\label{eq:epTransformTodisc}
\end{split}
\end{equation}
With equations (\ref{eq:energyIndisc}), (\ref{eq:pIndisc}) and (\ref{eq:epTransformTodisc}), $\hat{\textbf{S}}'$ can be expressed as 
\begin{equation}
    \hat{\textbf{S}}' = \frac{1}{w}\left\{\hat{\textbf{S}}+\left[\frac{\gamma-1}{\textbf{v}^2}\left(\textbf{v}\cdot\hat{\textbf{S}}\right)-\gamma \right]\textbf{v}\right\} ,\label{eq:shatdisc}
\end{equation}
where
\begin{equation}
    w = \gamma\left(1-\textbf{v}\cdot\hat{\textbf{S}}\right).
\end{equation}

The energy per photon, $\epsilon$, transforms analogous to the incident radiation energy (eq. \ref{eq:epTransformTodisc}), $\epsilon' = \epsilon w$. The photon number density, $n$, is part of the 4-current vector $(n,n\hat{\textbf{S}})$, so it also transforms as  $n' = n w$. The initial energy can thus be rewritten as 
\begin{equation}
    E_i = \frac{E_i'}{w} = \frac{n' A' \epsilon'}{w} =wn\epsilon \sigma_T'.\label{eq:initEnergyNS}
\end{equation}

We assume that the disc particle completely absorbes the radiation and reemits it isotropically in its own frame. The outgoing momentum $\textbf{p}_o'$ is thus zero and the outgoing energy $E_o'=E_i'$ in the particle frame. The outgoing 4-vector $(E_o',\textbf{0})$ in the neutron star frame is (using eq. \ref{eq:epTransformTodisc})
\begin{equation}\begin{split}
    E_o &= E_i' \gamma = \left[E_i \gamma \left(1-\textbf{v}\cdot\hat{\textbf{S}}\right)\right]\gamma\\
    \textbf{p}_o &= E_i' \gamma \textbf{v} = \left[E_i \gamma \left(1-\textbf{v}\cdot\hat{\textbf{S}}\right)\right]\gamma \textbf{v}.\label{eq:epfinalNS}
\end{split}
\end{equation}

The change in momentum of the radiation per unit time is $\textbf{p}_o-\textbf{p}_i$. The momentum exchange exists only between the radiation and the disc particle. For the disc particle, the change in momentum $\textbf{p}_p$ is (using eq. \ref{eq:initEnergyNS} and \ref{eq:epfinalNS})
\begin{align}
    \begin{split}
        \frac{d\textbf{p}_p}{d\tau} &= \textbf{p}_i-\textbf{p}_o=E_i\hat{\textbf{S}} - E_i\gamma^2\textbf{v}\left(1-\textbf{v}\cdot\hat{\textbf{S}}\right) \\
        &= E_i\left(\hat{\textbf{S}} - \gamma\textbf{v}w\right) = wn\epsilon \sigma_T'\left(\hat{\textbf{S}} - \gamma\textbf{v}w\right).
    \end{split}
\end{align}
Since $d/d\tau = \gamma\,d/dt$, the 3-force, $\textbf{F}_p$, exerted on the disc particle is 
\begin{equation}
    \textbf{F}_p = \frac{d\textbf{p}_p}{dt} = \gamma^{-1} wn\epsilon \sigma_T'\left(\hat{\textbf{S}} - \gamma\textbf{v}w\right).\label{eq:forcePR}
\end{equation}
For the $n\epsilon$ term in eq. (\ref{eq:forcePR}), we use the intensity parameter \citep{Bakala2019}
\begin{equation}
    \Phi^2 = \frac{L}{4\pi \sqrt{R_{b,q}(r)}}.
\end{equation}

\begin{figure}
    \centering
    \includegraphics[width = 0.45 \textwidth]{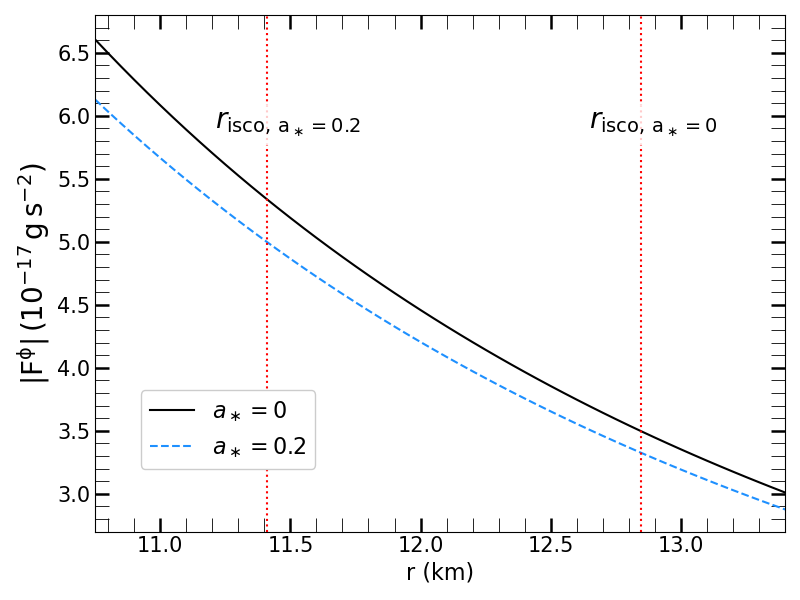}
    \caption{Angular force in the equatorial plane due to PR drag at the peak of the burst (eq. \ref{eq:forcePR}). The $a_*=0$ disc (black solid line) loses more angular momentum at a specific radius than the $a_*=0.2$ disc (blue dashed line). However, because the $a_*=0.2$ disc starts closer to the neutron star (left red dotted line) than the $a_*=0$ disc (right red dotted line), it experiences a higher flux, which offsets the impact of spin.}
    \label{fig:fphi}
\end{figure}

Fig. \ref{fig:fphi} shows the magnitude of the angular component of $\textbf{F}_p$ (eq. \ref{eq:forcePR}), $F^\phi$, at the peak luminosity $10^{38}$ ergs s\textsuperscript{-1}. Comparing the angular force at the same radius gives the expected result that $F^\phi$ is stronger if $a_*=0$ (solid black line) than if $a_*=0.2$ (blue dashed line). However, the $a_*=0.2$ disc is closer to the neutron star (left red dotted line) than the $a_*=0$ disc (right red dotted line), offsetting the impact of spin. The angular force is stronger by a factor of $\approx1.43$ for the $a_*=0.2$ disc than for the $a_*=0$ one at their respective $r_{\mathrm{isco}}$. This factor approximately matches the difference in inflowing mass accretion rate enhancement of the spin versus the no-spin simulation. Thus, the distance depencence of the PR drag can explain why the mass accretion rate in the spin simulation is more enhanced than in the no-spin simulation. 
The closer distance of the disc to the neutron star in the $a_*=0.2$ simulation outweighs the reduction of PR drag by spin, so that the disc in the spin simulation experiences a stronger PR drag than in the no-spin simulation. 

\section{Discussion}\label{sec:discussion}

In the burst simulations presented in Section \ref{sec:results}, increased PR drag removes angular momentum from the accretion disc (Fig. \ref{fig:omegaBurst}, \ref{fig:omegaRatio}), which increases the mass accretion rate during the burst (Fig. \ref{fig:mdotBurst}, \ref{fig:mdotTimeR}). An enhanced mass accretion rate is a potential explanation for an increase in persistent emission during observed X-ray bursts, which can be quantified via a normalization factor, $f_a$ \citep[e.g.,][]{Worpel2013,Worpel2015}. The normalization factors obtained from fitting observational data depend on the burst strength. They tend to peak at $\lesssim 4$ for non-photospheric radius expansion bursts \citep[e.g.,][]{Worpel2015,Keek2018,Guver2022MNRAS,Guver2022}, but can reach values $\gtrsim 10$ for photospheric radius expansion bursts \citep[e.g.,][]{Chen2019,Jaisawal2019,Roy2021}. However, how the normalization factor translates into changes in $\dot{M}$ remains unclear. Hence, the inflowing mass accretion rate enhancements of $\gtrsim8$ at the respective $r_{\mathrm{isco}}$ of the burst simulations (Fig. \ref{fig:mdotTimeR}) are quite plausible.

The normalization factor has also been used to account for a soft excess in the X-ray spectra of bursts \citep[e.g.,][]{Keek2018, Bult2019, Guver2022MNRAS, Guver2022}. As Fig. \ref{fig:mdotBurst} and \ref{fig:mdotTimeR} show, the inflowing mass accretion rate is significantly enhanced in spin and no-spin environments and could therefore contribute to the soft excess. However, another contribution to the soft excess may come from the reflection signature of the accretion disc \citep[e.g.,][]{Speicher2022}, which is disc temperature dependent. While moderate disc heating increases the soft excess, too much heating can move the soft X-ray emission to higher energies and thus lower the soft excess. Given warmer steady-state temperatures (right panel, upper row in Fig. \ref{fig:tempAll}), the $a_*=0.2$ disc is hotter than the $a_*=0$ disc during the burst (Fig. \ref{fig:tempAll}). A highly spinning star with a correspondingly hot disc may therefore emit low amounts of soft excess through reflection. Combined with an increased PR drag due to an accretion disc closer to the neutron star, the relative importance of PR drag in explaining the soft excess compared to reflection could increase with increasing neutron star spin.

\section{Conclusions}\label{sec:conclusion}
X-ray burst observations often feature an increase in persistent emission, which is usually quantified by a normalization factor, $f_a$ \citep[e.g.,][]{Worpel2013,Worpel2015,Jaisawal2019,Guver2022}. Moreover, $f_a$ has been used to explain a soft excess in the spectra of bursts \citep[e.g.,][]{Keek2018,Bult2019,Guver2022MNRAS,Guver2022}. The normalization factor is often linked to an increased mass accretion rate due to increased PR drag. While studies by \cite{Walker1989} and \cite{Fragile2020} showed that PR drag can significantly enhance $\Dot{M}$, \cite{Walker1992} alluded to a reduced impact of PR drag in the presence of neutron star spin.

In this paper, we studied the role of spin on PR drag during X-ray bursts by simulating thin accretion discs in non-spinning and spinning neutron star environments. We showed that while spin reduces the PR drag at a given radius, since the disc around a spinning star reaches closer to the surface, it actually experiences a stronger radiative flux, which offsets the PR drage reduction due to spin (Section \ref{subsec:angularForce}). Thus in both the no-spin and the spin simulations, PR drag is strong enough to remove angular momentum from the disc, which enhances the mass accretion rate and drains the inner disc region (section \ref{sec:results}). PR drag therefore remains a viable explanation for the increased persistent emission and soft excess during X-ray bursts. In addition, the X-ray burst heats the disc in the spin simulation to higher final temperatures than in the no-spin one due to different steady-state structures, which could lower the importance of reflection in contributing to the soft excess for highly spinning neutron star systems (Section \ref{sec:discussion}). 

\section*{Acknowledgements}
P.C. Fragile acknowledges support from National Science Foundation grants AST-1907850 and PHY-1748958. J. Speicher acknowledges support from the NDSEG fellowship.

\section*{Data Availability}
The data underlying this article will be shared on reasonable request to the corresponding author.

%

\vspace{5mm}



\bibliographystyle{mnras}
\bibliography{references}{}



\bsp	
\label{lastpage}
\end{document}